\documentclass{chi2008pubsformat}
\usepackage{times}
\usepackage{graphicx}

\title{
  AceWiki: A Natural and Expressive Semantic Wiki
}

\numberofauthors{1}

\author{
  \alignauthor Tobias Kuhn\\
    \affaddr{Department of Informatics}\\
    \affaddr{University of Zurich}\\
    \email{tkuhn@ifi.uzh.ch}
}

\begin{document}

\maketitle

\abstract{
We present AceWiki, a prototype of a new kind of semantic wiki using the controlled natural language Attempto Controlled English (ACE) for representing its content. ACE is a subset of English with a restricted grammar and a formal semantics. The use of ACE has two important advantages over existing semantic wikis. First, we can improve the usability and achieve a shallow learning curve. Second, ACE is more expressive than the formal languages of existing semantic wikis. Our evaluation shows that people who are not familiar with the formal foundations of the Semantic Web are able to deal with AceWiki after a very short learning phase and without the help of an expert.
}

\keywords{Controlled Natural Language, Attempto Controlled English (ACE), Semantic Wiki, Semantic Web, Ontology, Usability Test, Experiment}

\classification{H.5.2 User Interfaces; I.2.4 Knowledge Representation Formalisms and Methods.}

\section{Introduction}

Ontologies are mostly defined within communities. Semantic wikis \cite{tazzoli04semwiki} could support the community processes for building and maintaining an ontology. Unfortunately, existing semantic wikis are often hard to understand for novices and do not have sufficient support for expressive ontology languages.

\begin{figure}[tb]
  \begin{center}
    \includegraphics[width=8.3cm]{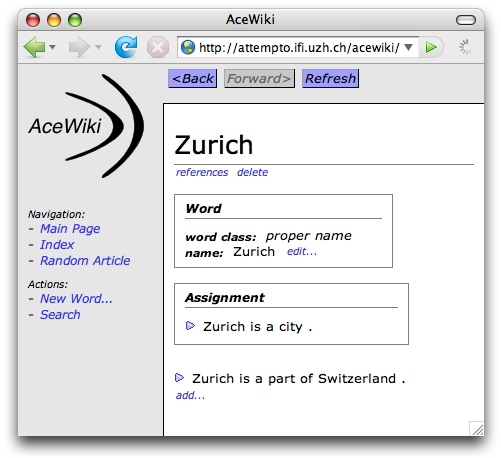}
    \caption{The web interface of the AceWiki prototype.}
    \label{fig:window}
  \end{center}
\end{figure}

AceWiki\footnote{\texttt{http://attempto.ifi.uzh.ch/acewiki}} is a prototype that tries to solve both problems and demonstrates the use of controlled natural language for semantic wikis. Being a semantic wiki, it combines the ideas and technologies of the Semantic Web with the concepts of wikis. The use of controlled natural language allows ordinary users --- who are not familiar with the concepts of logic and ontologies --- to understand, modify, and extend the formal content of the wiki. Figure \ref{fig:window} shows a screenshot of the AceWiki prototype.

\section{Background}

In this section, we refer to existing semantic wikis and explain their concepts. Next, we introduce Attempto Controlled English (ACE) which is the controlled natural language that is used for AceWiki.

\subsection{Semantic Wikis}

\textit{Platypus Wiki} \cite{campanini04platypus} was one of the first semantic wikis. It was introduced in 2004 and influenced many of the subsequent systems. The formal statements are represented in RDF and Platypus Wiki does not try to hide the technical details of this language. Thus, it is aimed at users who have experience with RDF. \textit{WikSAR} \cite{aumueller05towards} is another semantic wiki that has a similar design, but in contrast it does not show the RDF-specific details and has a very simple interface. \textit{SemperWiki} \cite{oren05semperwiki} is a third example following a similar approach. All those three systems are very RDF-centric.

\textit{Rhizome} \cite{souzis05rhizome} belongs to another category of semantic wikis, making use of RDF as well. Its goal is to ``represent informal, human-authored content in a semantically rich manner''. It allows the users to organize the informal knowledge with the help of formal structures. The focus is rather on management of knowledge than on advanced reasoning.

\textit{IkeWiki} \cite{schaffert06ikewiki} is a rich and sophisticated semantic wiki. It has a special focus on reusing existing informal articles (e.g.\ from Wikipedia) and to augment them with formal annotations. \textit{Semantic MediaWiki} \cite{kroetzsch07semwikipedia} is probably the most popular and most mature semantic wiki. It relies on the same wiki engine as the well-known Wikipedia and the goal of the authors is even to integrate their extended wiki syntax into Wikipedia. In both cases, RDF and OWL are used in the background.

\textit{SweetWiki} \cite{buffa06sweetwiki} is another semantic wiki approach. It is intended specifically for clearly identified communities, e.g.\ in intranets.

All those semantic wikis have in common that they rely on RDF and support only simple \textit{subject-predicate-object} structures. Some of them make use of the more expressive language OWL, but only to define the background model in a very static way. For example, none of the wikis allows the users to define general concept inclusion axioms like ``every person who writes something is an author''. For many applications simple RDF-triples may be sufficient, but there are cases where more expressive languages are needed. For example, there are several existing ontologies that exploit the expressivity of OWL, e.g.\ GALEN\footnote{\texttt{http://www.co-ode.org/galen/}} and the Ordnance Survey Hydrology ontology\footnote{\texttt{http://www.ordnancesurvey.co.uk/ontology/}}. It would be very convenient if these ontologies could be managed collaboratively within a wiki.

It is interesting that all the wikis addressed above organize their formal relations in ``annotations'' or as ``metadata''. This indicates that the formal statements are not considered the main content but rather an enrichment thereof.

In contrast, the approach of the \textit{myOntology} project \cite{siorpaes07myontology} is to build and maintain ontologies within a dedicated wiki. But again, it focuses on lightweight (i.e.\ relatively inexpressive) ontologies. They claim that ``... most users can not be expected to be able to add axioms''. We show that this is not true if controlled natural language is used.

\subsection{Attempto Controlled English}

AceWiki uses the controlled natural language Attempto Controlled English (ACE) \cite{fuchs:flairs2006}. ACE looks like English but avoids the ambiguities of natural language by restricting the syntax \cite{aceconstruction} and by defining a small set of interpretation rules \cite{aceinterpretation}. The ACE parser\footnote{\texttt{http://attempto.ifi.uzh.ch/ape}} translates ACE texts automatically into \textit{Discourse Representation Structures} \cite{fuchs08drs} which are a syntactical variant of first-order logic. Thus, every ACE text has a single and well-defined formal meaning. ACE supports a wide range of natural language constructs, e.g.\ singular and plural noun phrases, active and passive voice, relative phrases, anaphoric references, existential and universal quantifiers, negation, and modality. (However, for AceWiki we use only a subset of ACE and support not all of those constructs.) ACE has successfully been applied for different tasks, e.g.\ as a query language for ontologies \cite{bernstein04talking}, as a knowledge representation language for the biomedical domain \cite{kuhn2006protein}, and as a rule language for a multi-semantics rule engine \cite{kuhn07acerules}.

Furthermore, ACE has been used as a natural language front-end to OWL with a bidirectional mapping of ACE to OWL \cite{kaljurand07phd}. This mapping covers all of OWL 1.1 except data properties and some very complex class descriptions. AceWiki relies on this work for translating ACE sentences into OWL, which allows us then to do reasoning with existing OWL reasoners.

\section{Design}

The main goal of AceWiki is to improve knowledge aggregation and representation. AceWiki should be easier to use and understand than other semantic wikis. In addition, it should support a higher degree of expressivity. 

Unlike other semantic wikis, the formal statements are not contained in ``annotations'' and are not considered ``metadata'', but they are the main content of our wiki.

In order to achieve a good usability and still support a high degree of expressivity, AceWiki follows three design principles: \textit{naturalness}, \textit{uniformity}, and \textit{strict user guidance}.

By \textit{naturalness} we mean that the formal semantics has a direct connection to natural language. \textit{Uniformity} means that only one language is used at the user-interface level. \textit{Strict user guidance}, finally, means that a predictive editor ensures that only well-formed statements are created by the user. We will now discuss these three principles and show how they are achieved in AceWiki.

\subsection{Naturalness}

AceWiki is natural in the sense that the ontology is represented in a form that is very close to natural language. This requires a direct mapping of ontological entities to natural language words. In AceWiki, individuals are represented as proper names (e.g.\ ``Switzerland''), concepts\footnote{in OWL called \textit{classes}} are represented as nouns (e.g.\ ``country''), and roles\footnote{in OWL called \textit{properties}} are represented as transitive verbs (e.g.\ ``overlaps-with'') or as of-constructs (e.g.\ ``part of''). Using those words together with the predefined function words of ACE (e.g.\ ``every'', ``if'', ``then'', ``something'', ``and'', ``or'', ``not''), we can express ontological statements as ACE sentences. Since every ACE sentence is a valid English sentence, those ontological statements can be immediately understood by any English speaker.

We believe that ontological terms like ``property'', ``range'', or ``subclass'' are unknown or unclear to most potential users of a semantic wiki. Such terms do not comply with our principle of naturalness. We show that it is possible to avoid them, not only for the knowledge representation itself but also for captions, labels, help pages, etc. For example, instead of saying something like ``man is a subclass of human'' that uses the ontological term ``subclass'', we can simply say ``every man is a human'' which does not use any special terms. In the cases where there is no such solution, we use linguistic terms like ``noun'', ``verb phrase'', or ``sentence'' instead of ontological terms. Those linguistic terms should be familiar to most users, since they are taught even in elementary schools.

A minor problem arises when using controlled natural language. Since informal (uncontrolled) natural language is still needed sometimes (e.g.\ for labels, help pages, introductory notes, etc.), we have to make sure that the user does not confuse informal natural language with ACE. For example, an informal introductory note could be misinterpreted as a formal statement, or a formal statement could be misinterpreted as an informal explanation. In order to overcome this problem, we use a very simple convention: normal font is used for formal statements and terms, whereas italics are used for informal statements and terms in uncontrolled language. In this way, a user can immediately find out whether a certain statement or term is part of the formal ontology or not.

\begin{figure}[tb]
  \begin{center}
    \includegraphics[width=6.5cm]{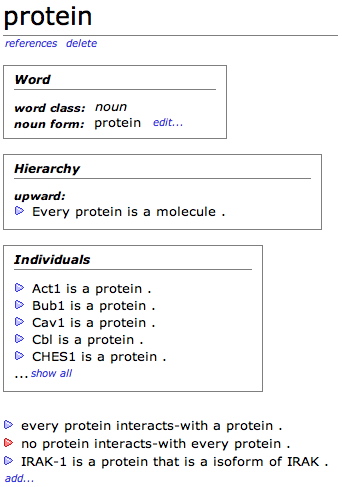}
    \caption{This is a screenshot of an article of an exemplary AceWiki instance. The title denotes the topic of this article, i.e.\ the concept ``protein''. The first box shows the linguistic information of this entity. The other two boxes contain sentences that follow a certain pattern, i.e.\ hierarchy statements and individual assignments. The unrestricted sentences are at the bottom of the page. All the text that is not in italics is ACE. The blue triangle in front of an ACE sentence denotes that this sentence is within OWL. Otherwise a red triangle is shown.}
    \label{fig:article}
  \end{center}
\end{figure}
Figure \ref{fig:article} shows a screenshot of an example wiki about proteins. All the formal representations appear in ACE, and all the text that is not ACE is displayed in italics.

\subsection{Uniformity}

The Semantic Web community defines three categories of languages on the logic level of the Semantic Web stack: ontology languages (e.g.\ OWL), rule languages (e.g.\ SWRL), and query languages (e.g.\ SPARQL). Most languages cover only one of those categories, and languages of different categories look usually very different.

This distinction makes sense from a knowledge engineer's point of view: Ontology languages need a different kind of reasoning than rule languages, and query languages ask for knowledge whereas ontology and rule languages assert knowledge. But this should not lead us to the conclusion that the end-users have to learn three different languages.

We claim that at the user-interface level ideally one single language should cover all those categories. In the background, there might be several internal languages, but the users should need to learn only one. For many users who are not familiar with formal conceptualizations, learning one formal language is already a hard task. We should not make this learning effort harder than necessary.

ACE is able to represent those different kinds of formal statements in a very natural way. In the case of queries, this distinction does not need to be made explicit: If a sentence ends with a question mark then it is clear for the user that this is a query and not an assertion. However, queries are still future work for AceWiki.

AceWiki classifies declarative ACE sentences into three categories: Some can be translated into OWL, others can be translated into SWRL, and finally there are ACE sentences that have no representation in OWL or SWRL at all. In ACE, this distinction is not visible and we think that users should not bother about it. The only thing they need to know is that if using an OWL reasoner only the OWL-compliant sentences are considered.

For that reason, AceWiki marks statements that are within OWL with a blue triangle, and all the other statements with a red triangle (as it can be seen on Figure \ref{fig:article}). Statements that are within SWRL could be marked in a similar way.

\subsection{Strict User Guidance}

Learning a new formal language is normally accompanied by frequent syntax error messages from the parser. Wikis are supposed to enable easy and quick modifications of the content, and syntax errors can certainly be a major hindrance in this respect, especially for new users.

This problem can be solved by guiding the users during the creation of new statements in a strict manner. By strict we mean that the creation of syntactically incorrect sentences is simply made impossible. This can be achieved by a predictive editor that guides the user step by step and ensures the syntactic correctness. Syntactic correctness can be subdivided into lexical correctness and grammatical correctness. Lexical correctness means that only the words that are defined in a certain lexicon are used. Grammatical correctness means that the grammar rules are respected.

To some degree, a predictive editor could also take care of the semantic correctness. It could prevent the users from adding statements that introduce inconsistency into an underlying ontology. If the verb ``meets'', for example, is defined in the ontology as a relation between humans then the predictive editor could prevent the user from writing sentences like ``a man meets a car'', assuming that the ontology says that ``car'' is not human.

AceWiki has a predictive editor that is used for the creation and modification of ACE sentences. It ensures lexical and grammatical correctness of the resulting sentences. The semantic correctness is not enforced, but the words that seem to be semantically suitable are shown first in the list. The suitable words are retrieved on the basis of the hierarchy of concepts and roles and the domain and range restrictions of roles. For example, if a user creates the incomplete sentence ``Limmat flows-through'' and there is a range restriction that says ``if something flows-through something Y then Y is a city'' then the individuals that are known to be cities are shown first in the list.

\begin{figure*}[tb]
  \begin{center}
    \includegraphics[width=12.5cm]{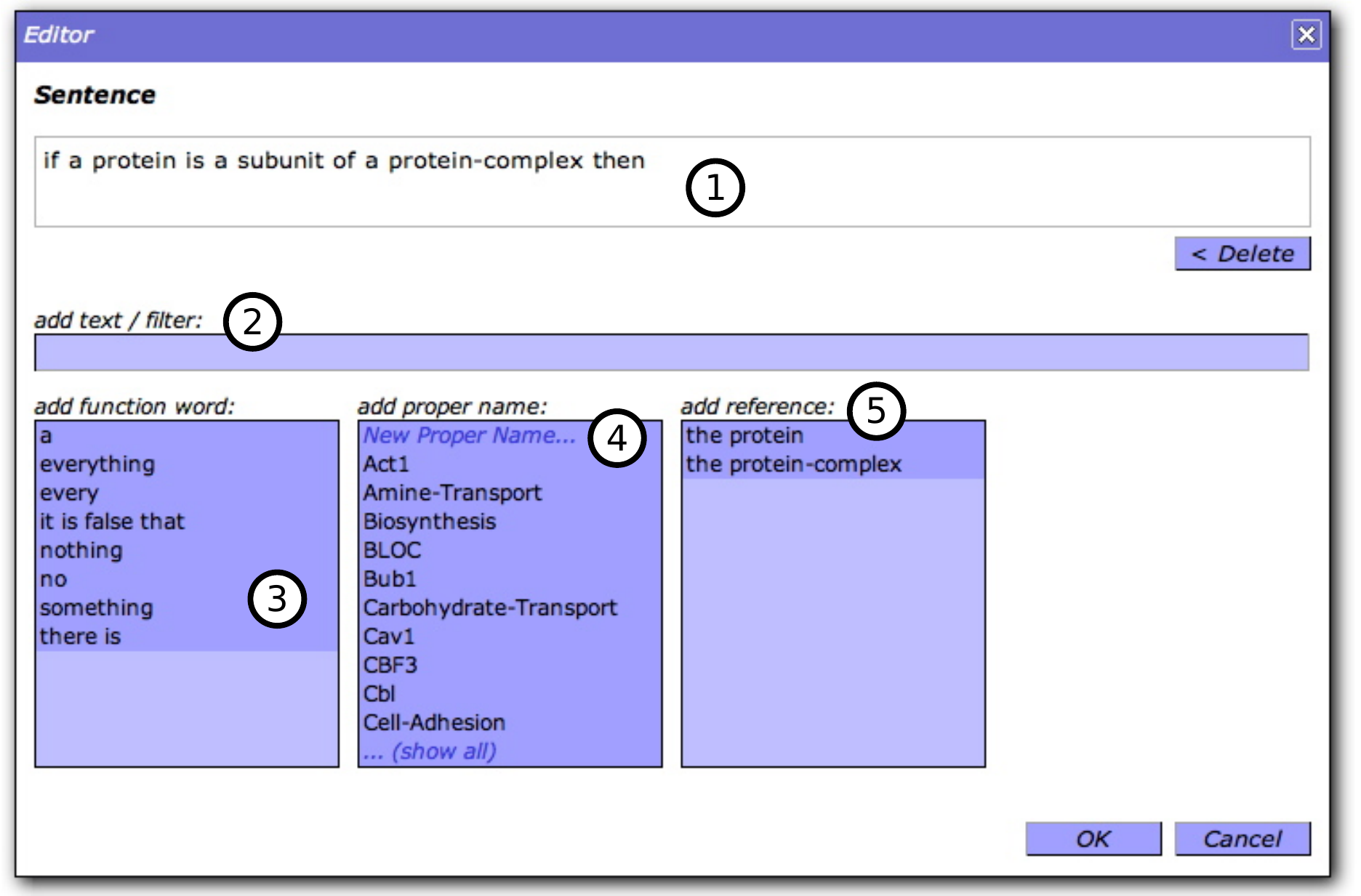}
    \caption{This is a screenshot of the predictive editor of AceWiki. (1) is a read-only text field that shows the beginning of an ACE sentence. This beginning has been entered by the user and it has been accepted by the predictive editor as a correct sentence beginning. The button ``Delete'' can be used to undo the last step. The text field (2) is used for entering the next words of the sentence. If they are accepted then they are moved to the text field (1). The tab key triggers autocompletion. (2) can also be used to filter the entries of the menus (3). Clicking on the entries of the menu boxes (3) is an alternative way to construct a sentence. There is a menu box for each word class that is allowed at the current position. If a word is not yet known then it can be added on-the-fly by clicking on the respective menu entry (4). Furthermore, references can be introduced that point to objects occurring earlier in the sentence (5).}
    \label{fig:editor}
  \end{center}
\end{figure*}

In order to be convenient for both novices and advanced users, the stepwise creation of a sentence can be done either by clicking on lists of proposed words (for novices) or by typing the words in a text field (for advanced users). Both alternatives are supported by a single graphical interface allowing the users to switch from one to the other at any time. Figure \ref{fig:editor} shows a screenshot of the predictive editor of AceWiki.

AceWiki provides special support for often used sentence patterns, i.e.\ concept and role hierarchies, domain and range of roles, and individual assignments. Such sentences are kept in separate boxes in the articles. The same predictive editor is used to create and modify them but with a reduced grammar that covers just the respective subset of ACE.

\section{Evaluation}

In order to evaluate AceWiki, we set up a small-scale end-user experiment. The hypothesis to be tested was whether average people (i.e.\ people who are not familiar with ontologies and logic) are able to learn how to deal with AceWiki within a short amount of time and without the help of an expert.

\subsection{Experiment Design}

The experiment was performed through the internet and it had a very simple design. The prerequisites for participation were only basic English skills and access to a computer with a broadband internet connection. We recruited 20 participants.

The participants received an instruction sheet which they read before they started with the experiment. These instructions explained the procedure and the task, but they did not explain how to interact with the AceWiki interface.

After reading the instructions, the participants were ready to start with the experiment. In the end, they filled out a questionnaire which asked for their background and their experiences with AceWiki.

The task for the participants was to visit AceWiki and to add knowledge to it. They were free to choose what kind of knowledge to add, as long as they followed three restrictions:
\begin{itemize}
\item The participants should add only knowledge that is true or at least true in most cases.
\item The knowledge should be general, i.e.\ verifiable by others.
\item The participants were allowed and encouraged to change or even delete the contributions of other participants if they found them violating one of the first two restrictions.
\end{itemize}
Furthermore, the participants were encouraged to add a couple of complex sentences starting with ``every'', ``no'', or ``if''.

Altogether, each participant should spend between 30 minutes and two hours (possibly split into several intervals) with\-in a time-frame of 14 days.

\subsection{Results}

The basis for the evaluation of the experiment were the questionnaire and the log files from the server.

Most of the 20 participants were students or graduates. Two participants had a computer science background, but they were not experts in the fields of Semantic Web or logic. The table below shows the exact distribution.
\begin{center}
\tabcolsep0.1cm
\begin{tabular}{l|rr}
  participants in total & 20 & \textit{(100\%)} \tabularnewline
  \hline
  students in computer science & 1 & \textit{(5\%)} \tabularnewline
  graduates in computer science & 1 & \textit{(5\%)} \tabularnewline
  other students & 8 & \textit{(40\%)} \tabularnewline
  other graduates & 8 & \textit{(40\%)} \tabularnewline
  no academic background & 2 & \textit{(10\%)} \tabularnewline
\end{tabular}
\end{center}

\begin{figure*}[tb]
  \begin{center}
    \includegraphics[width=13.5cm]{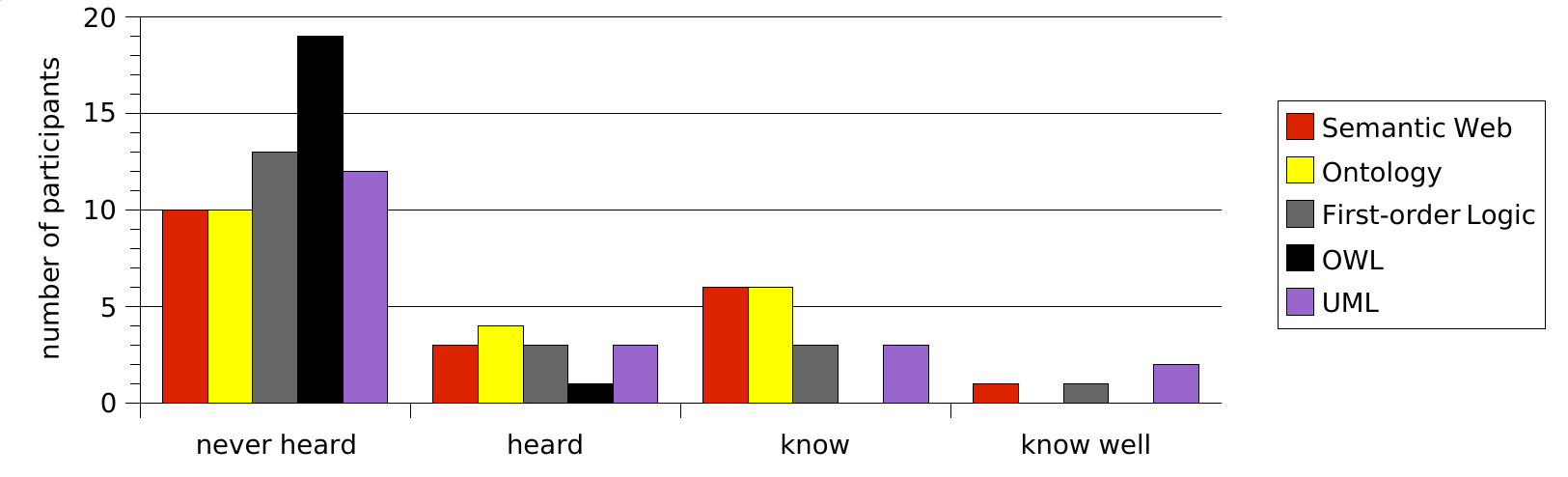}
    \caption{This chart shows the familiarity of the participants with the terms \textit{Semantic Web}, \textit{ontology}, \textit{first-order logic}, \textit{OWL}, and \textit{UML}. This data was retrieved from five questions of the questionnaire. For each of the terms the question was ``How familiar are you with this term?'' and there were four choices: ``I have never heard it'', ``I have heard it but I don't really know what it means'', ``I know more or less its meaning'', ``I know this term (very) well''.}
    \label{fig:terms}
  \end{center}
\end{figure*}
In the questionnaire, the participants were asked how familiar they are with different technical terms. Figure \ref{fig:terms} shows the result. The term \textit{OWL} was almost completely unknown. The majority of the participants have never heard the terms \textit{first-order logic} or \textit{UML}. Only in the case of the terms \textit{Semantic Web} and \textit{ontology}, we have a substantial minority knowing those terms. The results show that the participants had no considerable background in Semantic Web technologies or similar fields.

\begin{figure*}[tb]
  \begin{center}
    \includegraphics[width=17cm]{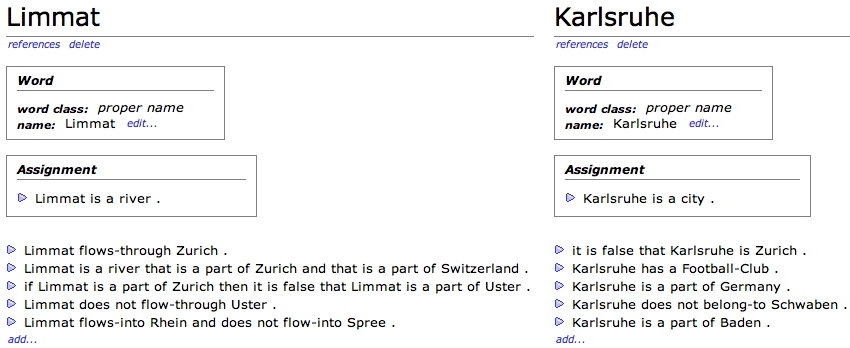}
    \caption{This figure shows the screenshots of two exemplary articles how they looked at the end of the experiment. All the sentences have been created by the participants of the experiment.}
    \label{fig:exparticles}
  \end{center}
\end{figure*}

Figure \ref{fig:exparticles} shows two examples of AceWiki articles how they came out of the experiment. Those screenshots intuitively show that the participants understood the ideas of AceWiki.

In order to measure how well the participants managed to work with AceWiki, we evaluated the sentences that they created or modified. For each of these sentences, we checked manually whether they represent a correct and sensible fact of the real world under the ACE semantics. The results are shown in the table below.
\begin{center}
\tabcolsep0.1cm
\begin{tabular}{l|rrrrr}
  & total & average & median & min & max\tabularnewline
  \hline
  $S$ & 186 & 9.3 & 7.0 & 1 & 31\tabularnewline
  $S^+$ & 148 & 7.4 & 6.0 & 1 & 22 \tabularnewline
  $S^-$ & 38 & 1.9 & 1.0 & 0 & 9 \tabularnewline
  $S^+/S$ & 0.796 & 0.854 & 0.912 & 0.5 & 1.0 \tabularnewline
\end{tabular}
\end{center}
The overall number of sentences $S$ is 186. 148 of them are considered correct and sensible ($S^+$), whereas the remaining 38 are not ($S^-$). We do not count the sentences that have been created and later removed by the same participant. If someone modified a sentence that was created by himself then we count only the last version. If a sentence was modified by another participant then the respective versions of the sentence count for each of the participants. Thus, this table shows the achievements of the individuals, not of the community.

Let us first explain how we judged whether a sentence is correct and sensible. The main criteria was that the sentence is a true statement (under the ACE semantics) of the real world using the common interpretations of the natural language words. If this is not the case, e.g.\ for ``every musician is a man'', then the sentence counted for $S^-$. In the case of 24\% of the incorrect sentences, words were used in the wrong category, for example ``every London is a city'' where ``London'' was introduced as a common noun instead of a proper name. Another 24\% of the incorrect sentences are statements like ``a city is a landscape-element'' which is interpreted in ACE as having only existential quantification: ``there is a city that is a landscape-element''. Even though this is a correct statement about the real world, the user probably wanted to say ``every city is a landscape-element''. For that reason, such sentences were considered incorrect. (The remaining 52\% of the incorrect sentences do not show specific patterns for further categorization.) On the other hand, sentences like ``every country is a part of a continent'' were considered correct, even though it depends on the interpretation of ``part of'' and ``continent''. One could say that Russia is not part of a continent, but only overlaps with Europe and Asia. But in this case, there is no reason to believe that the participant wanted to say something different than what the ACE semantics defines.

The results show that almost 80\% of the created sentences were correct, which is --- we think --- a very good result. Furthermore, every participant created at least one correct sentence. Another interesting result is that the ratio of correct sentences was in the worst case 50\%. Thus, no participant created more wrong sentences than correct ones. Altogether, we can conclude that all of the participants managed to deal with AceWiki.

Another interesting point to investigate is whether there was a fruitful community process. If we look at the results from a community perspective then we should not consider the sentences that have been removed later by someone else and we should count sentences that have been edited by different participants only once. Under these conditions, we get slightly different values.
\begin{center}
\tabcolsep0.1cm
\begin{tabular}{l|r}
  \hline
  $S_c$ & 179 \tabularnewline
  $S_c^+$ & 145 \tabularnewline
  $S_c^-$ & 34 \tabularnewline
  $S_c^+/S_c$ & 0.810 \tabularnewline
\end{tabular}
\end{center}
The ratio of correct sentences is slightly higher than the one from the individual data. This means that the community interaction --- i.e.\ the deletion and modification of other's sentences --- increased the quality of the content in this particular experiment. However, we do not have enough data for making a general statement.

\begin{figure*}[tb]
  \begin{center}
    \includegraphics[width=15cm]{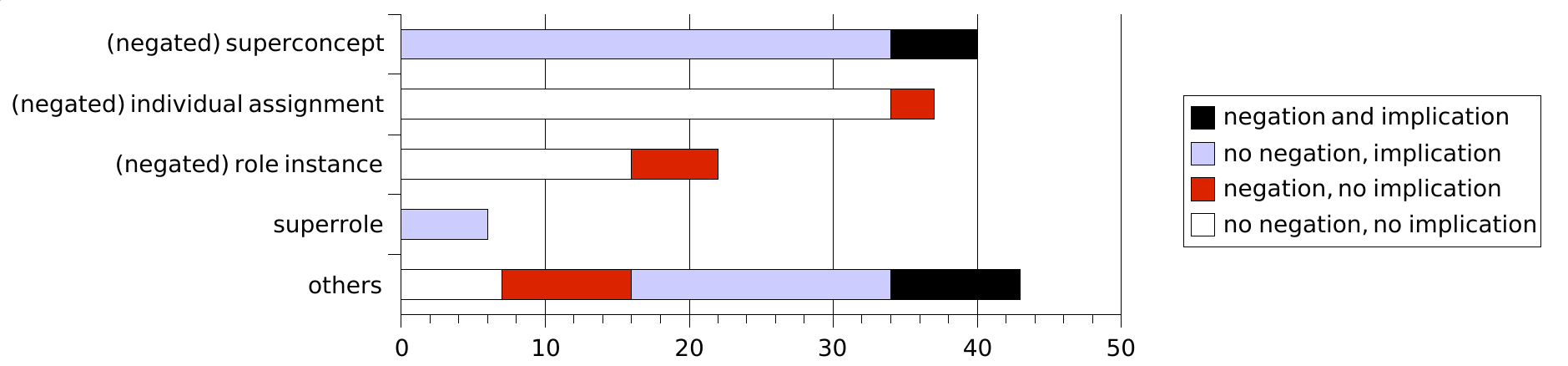}
    \caption{This chart shows the most frequent sentence patterns in absolute numbers. Only the correct sentences are considered.}
    \label{fig:types}
  \end{center}
\end{figure*}
The participants were encouraged to create not only simple sentences, but also some complicated ones. We can now find out whether they managed to do so. Figure \ref{fig:types} shows the most frequent sentence patterns and reveals the occurrence of negation (i.e.\ ``does not'', ``is not'', ``no'', or ``it is false that'') and implication (i.e.\ ``if ... then'', ``every'', or ``no''). Only the correct sentences are considered here.

The two most frequent sentence patterns were superconcept statements (positive e.g.\ ``every canal is a waterbody'' or ne\-gated e.g.\ ``it is false that every animal is a mammal'') and individual assignments (positive e.g.\ ``Zu\-rich is a city'' or ne\-gated e.g.\ ``Bob-Dylan is not a woman''). Also quite frequent were role instances (positive e.g.\ ``Limmat flows-through Zu\-rich'' or ne\-gated e.g.\ ``it is false that Winston-Churchill is a prime-minister of Denmark'') and superrole statements (e.g.\ ``if something X protects something Y then X shelters Y''). All the examples are sentences that the participants created during the experiment.

It is remarkable that there is a long tail of other sentence patterns and that 61\% of the sentences contained a negation or an implication or both. This shows that the participants made use of the high expressivity of ACE.

In order to evaluate the performance of the participants, we have to take the time dimension into account. The following table shows the time values (in minutes) that we retrieved from the log files.
\begin{center}
\tabcolsep0.1cm
\begin{tabular}{l|rrrrr}
  & total & average & median & min & max \tabularnewline
  \hline
  $t$ & 931.2 & 46.6 & 39.1 & 7.7 & 132.5 \tabularnewline
  $t_{f}$ & & 11.0 & 8.0 & 2.5 & 33.8 \tabularnewline
  $t/S^+$ & 6.3 & 8.2 & 7.2 & 2.9 & 24.7 \tabularnewline
\end{tabular}
\end{center}
The first line shows the overall time $t$ of the participants. This is the time they spent on AceWiki, not counting the time for reading the instructions and for filling out the questionnaire. The second line shows the time $t_f$ needed for creating the first correct sentence. The final line contains the time per correct sentence and shows how well the participants performed. Thus, they needed on average 11.0 minutes to create their first correct sentence, and overall the time per correct sentence was 8.2 minutes.

Those results do not look very spectacular at first sight, but we have to recall the situation of the participants. When we start counting the minutes, the participants see AceWiki for the very first time. The instructions contained no explanation whatsoever of the AceWiki interface. In order to get familiar with this unknown interface, the participants started to navigate around, searched for terms, and explored the predictive editor. Some of them added new words without adding a sentence yet, and some added a sentence but removed it again. And then, after only eleven minutes, on average, the participants managed to create their first correct sentence. Over the complete duration, they created a correct sentence approximately every eight minutes, and --- as Figure \ref{fig:types} shows --- most of those sentences were quite complicated. We think that these are very good results under the given circumstances, and they show that AceWiki has indeed a shallow learning curve.

Restricting our attention to the time when the editor window was open, we can evaluate the usability of the predictive editor alone. The following table shows these values.
\begin{center}
\tabcolsep0.1cm
\begin{tabular}{l|rrrrr}
  & total & average & median & min & max \tabularnewline
  \hline
  $t_{e}$ & 573.2 & 28.7 & 24.7 & 1.7 & 85.6 \tabularnewline
  $t_{ef}$ & & 5.3 & 3.0 & 0.9 & 28.6 \tabularnewline
  $t_{e}/S^+$ & 3.9 & 4.6 & 3.3 & 1.7 & 12.5 \tabularnewline
\end{tabular}
\end{center}
The editor window was on average open for 5.3 minutes (not necessarily continuously) before the first correct sentence was created, and overall the time per correct sentence was 4.6 minutes. Note that this includes also the time for wrong sentences, the time for sentences that have been canceled during their creation, and the time that was needed to introduce new words.

Again, we think that these are very good results considering that the users were not trained how to interact with the predictive editor.

\begin{figure}[tb]
  \begin{center}
    \includegraphics[width=8.0cm]{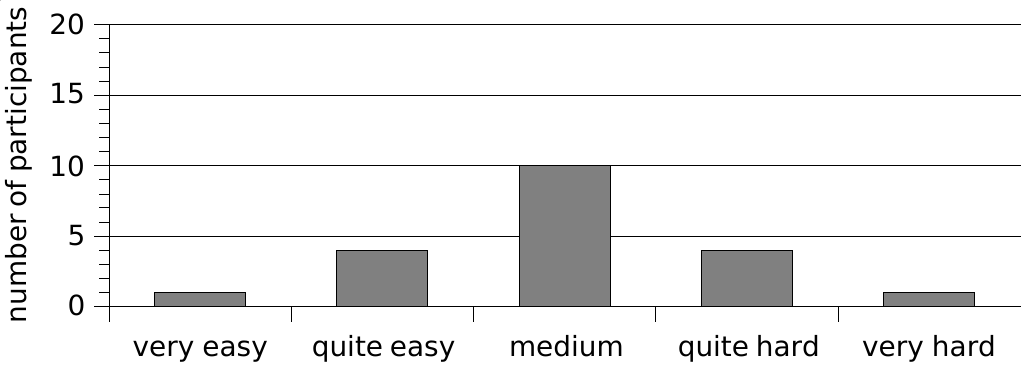}
    \caption{This chart shows how the participants answered the question ``How easy/difficult was the handling of AceWiki?'' after the experiment.}
    \label{fig:diff}
  \end{center}
\end{figure}

Finally, we can take a look at the participants' feedback after the experiment. In the questionnaire we asked how easy or how difficult the handling of AceWiki was. Figure \ref{fig:diff} shows the result. The responses are distributed symmetrically and have a peak at ``medium''. On the one hand, this is a good result since only 25\% of the users found it hard to use AceWiki. We have to consider that the participants experienced only the costs of formal knowledge representation, but not the benefits (since reasoning features were missing). Furthermore, knowledge representation is inherently a difficult task. Probably, it will never be possible to make this very easy for everybody. On the other hand, the results show that there is certainly room for improvement.

\section{Future Work}

There are various possibilities to improve AceWiki. ACE supports a wide range of word classes and we plan to support more of them in AceWiki. Intransitive verb (e.g.\ ``flows'') or intransitive adjectives (e.g.\ ``large'') could be used to represent concepts. Transitive adjectives (e.g.\ ``loca\-ted-in'') and comparative adjectives (e.g.\ ``larger than'' or ``as large as'') could be used to represent roles. Furthermore, roles could be represented by adverbs together with transitive verbs (e.g.\ ``directly interacts-with'').

Another important extension would be to support number restrictions like ``every car has exactly 4 wheels'' or ``every person has at most 2 parents''. Such sentences are already supported by ACE and also by the ACE-to-OWL translator.

In order to enable reasoning with the formal content of Ace\-Wiki, we plan to tightly integrate an existing OWL reasoner. The interaction with the reasoner should be done via ACE questions, e.g.\ ``which river flows-through Zurich?'' or ``is Zurich a part of Europe?''. A very nice feature would be to support inline queries, as shown by other semantic wikis (e.g.\ WikSAR and IkeWiki).

AceWiki is still a prototype and it probably will stay so for the near future. It lacks a safe concurrency management, has only a rudimentary support for persistent storage, and it does not feature a history and undo facility. All those points would of course be crucial for a real-world application. So far, we focused on the aspects that are interesting from a scientific rather than an industrial point of view.

\section{Conclusions}

We presented the AceWiki prototype and an experimental evaluation of its usability. The fact that the participants of the experiment created on average a correct statement every eight minutes shows that AceWiki is easy to learn. Furthermore, the users were able to create complex statements that go beyond the expressivity of other semantic wikis. The results indicate that our three design principles --- naturalness, uniformity, and strict user guidance --- are beneficial for enabling unexperienced users to effectively interact with a semantic wiki.

Since AceWiki is still an emerging system and since we observed several possibilities for improvements during the experiment, we think that even better results will be possible in the future.

\section{Acknowledgement}

This research has been funded by the European Commission and the Swiss State Secretariat for Education and Research within the 6th~Framework Programme project REWERSE, and by the University of Zurich within the research grant program 2006 (Forschungs\-kredit). I would like to thank Norbert E. Fuchs and Kaarel Kaljurand for their inputs.

\end{document}